\documentclass[preprint,preprintnumbers,aps,prb]{revtex4}
\usepackage{graphicx}
\usepackage{bm}

\def\molsim#1#2#3{{ Mol. Simulat.} {\bf #1}, #2 (#3).}

\def\prl#1#2#3{{ Phys. Rev. Lett.} {\bf #1}, #2 (#3).}

\def\ibid#1#2#3{{ ibid.} {\bf #1}, #2 (#3).}
\def\pre#1#2#3{{  Phys. Rev. E.} {\bf #1}, #2 (#3).}

\def\prb#1#2#3{{ Phys. Rev. B} {\bf #1}, #2 (#3).}
\def\jcp#1#2#3{{ J. Chem. Phys.} {\bf #1}, #2 (#3).}

\def\jpcb#1#2#3{{ J. Phys. Chem. B} {\bf #1}, #2 (#3).}

\def\cpl#1#2#3{{ Chem. Phys. Lett.} {\bf #1}, #2 (#3).}

\def\mp#1#2#3{ { Mol. Phys.} {\bf #1}, #2 (#3).}

\def\science#1#2#3{{ Science} {\bf #1}, #2 (#3).}
\def\nature#1#2#3{ { Nature} {\bf #1}, #2 (#3).}

\def\jpcm#1#2#3{ { J. Phys.: Condensed Matter } {\bf #1}, #2 (#3).}

\def\pccp#1#2#3{ { Phys. Chem. Chem. Phys.} {\bf #1}, #2 (#3).}
\def\jncs#1#2#3{ { J. Non-Cryst. Soilds} {\bf #1}, #2 (#3).}

\def\degc{$^\circ$C}

\def\onebyf{$1/f^\alpha$}

\def\expt#1{\langle #1\rangle}
\def\be{\begin{equation}}
\def\ee{\end{equation}}

\def\br{{\bf r}}

\begin{document}
\baselineskip 20pt
\begin{center}
{\Large \bf Diffusional anomaly and network dynamics in liquid silica}\\
\qquad \\
{\bf Ruchi Sharma}, {\bf Anirban Mudi} and
{\bf Charusita  Chakravarty}$^*$\\
Department of Chemistry,\\
Indian Institute of Technology-Delhi,\\
New Delhi: 110016, India.\\
 \quad\\
 {\bf Abstract}
 \end{center}
The present study applies the power spectral analysis technique to understand
 the diffusional anomaly in liquid silica, modeled using the BKS potential. 
Molecular dynamics simulations have been carried out to show that power spectrum
of tagged particle potential energy of silica shows a regime with 
$1/f^{\alpha}$ dependence on frequency $f$ which is the characteristic 
signature of multiple time-scale behaviour in networks. As demonstrated earlier in the case
of water (J. Chem. Phys., {\bf 122}, 104507 (2005)), the variations in the 
mobility associated with the diffusional anomaly are mirrored in the scaling
 exponent $\alpha$ associated with this multiple time-scale behaviour. Our
results indicate that in the anomalous regime, as the local tetrahedral 
order decreases with temperature or pressure, the coupling of local modes 
to network reorganisations increases and so does the diffusivity. 
This symmetry-dependence of the vibrational couplings is responsible 
for the connection between the structural and diffusional anomalies.

\vfill
{* Author for correspondence (Tel: (+) 91 11 26591510; Fax: (+) 91 11 2686 2122; E-mail: {\tt charus@chemistry.iitd.ernet.in})\hfill}
\newpage
                                                                                
\section{Introduction}

Tetrahedral network-forming liquids are characterised by strong, local 
anisotropic interactions which impose a local tetrahedral order with four 
nearest neighbours, in contrast to the local
icosahedral order characteristic of the random closed-packing
structures seen in simple liquids. The interactions in simple
liquids are dominated by strong, short-range repulsions and 
weak isotropic attractions; typical examples are the rare gases and
liquid metals \cite{hm86}. The best known example of a tetrahedral
liquid is water where each oxygen atom can form four tetrahedrally
disposed hydrogen bonds with its neighbours. At room temperature, the
hydrogen bond energy is of the order of 5-10$k_BT$ and the hydrogen bonds
in water can break and reform on timescales of the order of picoseconds 
resulting in a fluctuating, random three-dimensional network of water 
molecules \cite{pgd03,mcbf}.
A second important example is liquid silica where the SiO$_4$ tetrahedra
are linked to form a random network \cite{ak76,abhst00}. Interactions in SiO$_2$ have a 
mixed ionic-covalent character with Si-O bond energy of approximately 450 kJ 
mol$^{-1}$ such that in the temperature regime above 2000K 
where the liquid phase is stable, it is of the order of 25$k_BT$ or 
less \cite{fam04}. 
 The liquid phases of Si, Ge, Sb, Bi and Ga
 are expected to be tetrahedral as are those of several ionic compounds such as 
BeF$_2$ and intermetallics such as InSb, GaAs, GaP, HgTe, CdTe and CdSe
\cite{ht02}.

The liquid state thermodynamic and kinetic 
properties of tetrahedral liquids are in many 
respects anomalous when compared with those of simple liquids 
\cite{pgd03,abhst00,rd96}.
The best known of the thermodynamic 
anomalies is the density anomaly shown by the existence of a temperature of
maximum density (TMD).  Unlike simple liquids where density decreases 
monotonically with temperature, in  tetrahedral liquids, increasing
thermal kinetic energy over certain density regimes
results in progressive destruction of the network, consequent
compaction and a resulting increase in density.  
At atmospheric pressure, the TMD
of water occurs at 279K whereas that of silica occurs at approximately
5000K. The behaviour of response functions of tetrahedral liquids, such as the 
isothermal
compressibility ($\kappa$) and the isobaric heat capacity ($C_p$), can also 
be anomalous.  In simple liquids , both $C_p$ and $\kappa$ decrease 
monotonically with decreasing temperature but, for tetrahedral liquids, both
$\kappa$ and $C_p$ as a function of temperature along an isobar  can show 
minima. Thermodynamic arguments can be used to show that the existence
of  a density anomaly implies the existence of compressibility and heat 
capacity anomalies \cite{pgd03}. In addition to the thermodynamic anomalies, 
tetrahedral liquids 
also show kinetic anomalies. For example, the diffusional anomaly corresponds
to a regime in which the diffusivity increases as a function of density,
in contrast to simple liquids where  diffusivity decreases with density due
to increasing steric hindrance.  Qualitatively, this
anomalous behaviour of the mobility can be understood in terms of the disruption
of the tetrahedral network by increasing pressure or
density   which facilitates translational motion of the water molecules.
The phase diagram of tetrahedral liquids can show a number of unusual
features, such as a negatively sloped melting line on the pressure-temperature
phase diagram.  There is also increasing evidence that tetrahedral liquids can 
show polyamorphism i.e. the existence of low and high-density liquid or glassy 
phases.

Over the past decade, a better understanding of the connection between the
various anomalous properties of tetrahedral liquids has been obtained
from a combination of experimental and simulation studies on water, and
to a lesser extent, silica.  A key innovation has been the introduction of
order metrics to quantify the nature as well as the extent of
structural order present in such liquids \cite{ed01,sdp02}. The extent of local
tetrahedral order present around a  tagged tetrahedral centre (e.g.O
atom in water and Si atom in SiO$_2$)  is gauged by
a tetrahedral order parameter $q$ defined as
\be
q = 1 -\frac{3}{8}\sum_{j=1}^3\sum_{k=j+1}^4 (\cos \psi_{jk}+ 1/3)^2
\ee
where $\psi_{jk}$ is the angle between the bond vectors {\bf r}$_{ij}$
and {\bf r}$_{ik}$ where $j$ and $k$ label the four nearest neighbour
atoms of the same type e.g. the 
$q_O$ order parameter in water  is  defined using the positions
of the four nearest oxygen atoms \cite{ed01,edt02}. 
The translational order parameter, $\tau$, was defined as
\be \tau = (1/\xi_c ) \int_0^{\xi_c} (g(\xi ) -1)d\xi \ee
where $\xi =r\rho^{1/3}$ and $\xi_c =2.83$ \cite{edt03}.
The translational order parameter, as defined above, will capture the extent of
pair correlations present in the liquid and can be thought of as a type of
density-ordering, since $\tau$ will increase as the random close-packing limit
is approached. The structure and dynamics of tetrahedral liquids can be 
understood in terms of an interplay between short-range tetrahedral ordering
and translational or density ordering which promotes local icosahedral order
seen in simple liquids. At a given temperature, $q$ will show a maximum and
$\tau$ will show a minimum as a function of density; the loci of these
extrema in the order define a structurally anomalous region in the
density-temperature ($\rho T$) plane where the tetrahedral and translational
order parameters are found to be strongly correlated. The region of the 
density anomaly,
where $\partial \rho /\partial T>0$, is bounded by the structurally
anomalous region. The diffusionally anomalous region ($\partial D /\partial\rho
>0$) encloses the boundaries of the structurally anomalous region
in the case of silica. In water, on the other hand, the boundaries
of the diffusionally anomalous region lie between those of the
structural and density anomalies. Thus one can think of tetrahedral
liquids as displaying  a cascade of anomalies where structural, diffusional 
and density anomalies occur consecutively as the degree of tetrahedral order
is increased. Our understanding of the molecular level processes which
give rise to this cascade effect is still incomplete. An important
step in this direction is the demonstration of the cooperative origin
of low-density domains in water which suggests that the probability
density of centres of high local tetrahedral order must reach a 
certain threshold value before cooperative effects can precipitate
the formation of ramified, low density clusters of water molecules
\cite{edt02,rlp04}.

In a series of recent papers, we showed that power spectral analysis could
be used to  understand the connections between the local order, hydrogen
bond network dynamics and the kinetic anomalies in water \cite{mrc03,mc041,mc042,mcr05,mc06,mcm06}.
The power spectral density of an observable $A(t)$ as a function of time $t$ is defined as follows \cite{numrec}: 
\be
\label{psd}
S(f) =\lim_{T\rightarrow\infty} \frac{1}{T} \left| \int_{-T/2}^{T/2} A(t)
e^{2{\pi}ift}dt\right|^2.
\ee
The power spectrum contains the same information as the
time correlation function of the observable but it highlights its behaviour 
over a range of frequencies. The multiple time-scale dynamics of networks
is reflected in the power spectrum, $S(f)$, as a $1/f^\alpha$ dependence
\cite{bm99,em02}.
In the case of water, where the network is formed by the coupling of
individual molecules by hydrogen bonds,  the observables which
are sensitive to local motions as well as network re-organisations
were found to be the local tetrahedral order parameter and the
tagged molecule potential energy. Fluctuations in these observables
generated power spectra with three key features: (a) a local vibrational
peak corresponding to the librational modes in water; (b) a multiple
time scale region with $1/f^\alpha$ behaviour and (c) a crossover to
white noise at lower frequencies identifying the time scales for decay
of correlations in the network. 
 The diffusivity was shown to be strongly 
correlated with the exponent $\alpha$ of the \onebyf\ region of the power 
spectra of the tagged particle potential energy fluctuations.  Thus
the power spectrum has been shown to provide a direct dynamical measure
in the scaling exponent $\alpha$   of the degree  of coupling of the 
librational modes to the network vibrations which complements
the static structural measures, such as the coordination number distributions,
used in earlier studies. Destruction of tetrahedral local order by
temperature, pressure or ionic solutes facilitates the coupling of local,
librational modes to the network motions and therefore the parallel behaviour
of the structural and diffusional anomalies is not surprising. The
power spectral analysis shows that the boundaries of the
two regions cannot be identical because the
structurally anomalous region is defined by the
average values of the local order parameters whereas the diffusionally anomalous
region is determined by the dynamical correlations in the fluctuations
in the local order and energy.

In this paper, we apply the power spectral analysis techniques described above
to understand the diffusional anomaly in liquid silica, modeled using the
BKS potential \cite{bks90,kfb91}.  While the behaviour
of liquid silica and water has been shown to be parallel in many respects, 
there are some important differences \cite{rd96,sdp02,vsgp04,vsp00,vps01,nss02}.
 Among all the crystalline polymorphs of silica, only $\beta$-cristobalite
has a melting line with a slightly negative slope having a change in 
 volume equal to -1\% as compared to -8.9\% seen for water \cite{fam04}.
From the chemical bonding perspective, silica is thought of as an ionic 
liquid with strong covalent character.  The tetrahedral motif in water
is displayed by the Walrafen-pentamer structure, while in silica
and many silicates, the SiO$_4$ tetrahedra are the crucial
structural unit. The O-H$\cdots$O hydrogen bond 
in water is nearly linear while in silica, the Si-O-Si bond angle varies
between 140-155\degc  \cite{ha97}.  
It is therefore interesting to consider the extent to which the
dynamical behaviour of the two networks is similar.
Based on our studies of water, we have found the tagged particle 
(atom or molecule) energies as a convenient dynamical variable
which is sensitive to local order and energy of the molecular environment.
In addition, since it is defined as the interaction of a tagged atom or
molecule with all other particles in the simulation cell, it is
reasonably sensitive to network reorganisations over a fairly
wide range of length and time scales. 
 Therefore we focus  on the tagged ion potential energies as the important 
local variable in a tetrahedral liquid and study the corresponding static 
distributions as well as the power spectrum of the fluctuations. 

The paper is organised as follows.
Section 2 discusses the computation of the tagged ion potential energies
when Ewald summation is used to evaluate the long range, Coulombic interactions.
Section 3 summarises the aspects of power spectral analysis required in
this work. Computational details regarding the potential model and molecular
dynamics simulations is given in Section 4. Section 5 and 6 contain 
the results and conclusions respectively.

\section{Computation of Tagged Molecule Potential Energies}

Parametric potentials for silica are typically constructed as a sum 
of long-range, Coulombic interactions, $U_{coul}$, and short-range
van der Waals interactions, $U_{vdw}$, such that 
\be U = U_{coul} + U_{vdw}\ee
The Coulombic contribution comes from the charge-carrying Si and O atoms.
We assume that the short-range 
interactions include only two-body terms, but if three-body terms are present
the computation can be adapted. The functional form of the
short-range potential used in this work is discussed in section 4.1.
For a pair additive system, if
$u_{Si}$ and $u_O$ are the tagged atom energies and $n_{Si}$ and $n_O$ 
are the number of Si and O atoms respectively, then the corresponding
ensemble averages are related to the average configurational energy
as: $2\expt{U}/N = (n_{Si}/N) \expt{u_{Si}} + (n_{O}/N)\expt{u_O}$.

The short range van der Waals interactions contribute an amount $(u_{i}^{vdw})$ 
to the tagged atom potential energy $u_i$. Since the
interactions are short-ranged, the minimum image convention can be
applied and $u_i^{vdw}$ can be evaluated by summing
over all pair interaction contributions between atom $i$ and 
all other atoms $j$ located in the central simulation cell.
If the short-range potential has a 
finite value at the spherical cut-off radius, a density-dependent long-range
correction term must be included in $u_i^{vdw}$.

To evaluate the Coulombic contribution to the tagged molecule potential
energy, one must consider the Ewald summation technique for evaluating
long-range forces \cite{at86,fs02}.
The charge density distribution for a collection of $N$ point charges located in
the central simulation cell is given by:
\be
\rho (\br ) = \sum_i^N q_i\delta (\br -\br_i)
\ee
Note that $\br$ is a 3-dimensional position vector and
$\br_i$ is the position of the $i^{th}$ charge $q_i$ and that
the central simulation cell is set up to be electrically neutral.
The simulation cell is taken as cubic with edge length $L$.
The Coulombic interaction energy between all charges in the central
simulation box and all other periodic images is given by:
\be \label{eq:eq5.42}
U_{coul}=\frac{1}{2}\sum_{{\bf n}=0}^{\infty}{'}\sum_{i=1}^{N}\sum_{j=1}^{N}\frac{{q_i}{q_j}}{4\pi\epsilon_{0}|{\bf r}_{ij}+{\bf n}|}
\ee
where ${\bf r}_{ij}$ is the vector distance between
charges $q_i$ and $q_j$, 
${\bf n} =(n_xL,n_yL,n_zL)$ denotes the set of possible translations
of the simulation cell. The central
simulation box corresponds to 
the ${\bf n}=0$ 
case and the
prime on the first summation indicates that the series does not include 
the interaction $i=j$ for ${\bf n}=0$. Unlike in the case of the short-range
van der Waals interactions, the inverse distance dependence of the
Coulombic interaction implies that the summation cannot be restricted to just the central simulation 
cell 
and the above series will converge very slowly.
Ewald summation replaces the slowly convergent series in 
equation (5) by a sum
of two rapidly convergent series obtained by replacing the
charge distribution $\rho (\br)$ by
a sum of two charge distributions, $\rho_{screen}$ and $\rho_{gauss}$.
To obtain $\rho_{screen}(\br )$, each point charge $q_i$ is surrounded
by a neutralizing charge distribution, usually a Gaussian distribution
centred at the location of $q_i$.  To ensure that 
$\rho(\br ) = \rho_{screen}(\br ) +\rho_{gauss}(\br )$, $\rho_{gauss}(\br )$ must
consist of Gaussian charges which neutralise the screening charges
used to construct $\rho_{screen}$. 

The potential energy contribution due to the screened charges is
denoted by $U_{screen}$. The advantage of choosing neutralising Gaussian
distributions of the form:
\be
{\rho_i}({\bf r})=\frac{{q_i}{\alpha}^3}{\pi^{3/2}}\exp(-{\alpha}^{2}{r^2})
\ee
is that $U_{screen}$ can be readily evaluated using 
the complementary error function, $erfc$, as follows:
\be \label{eq:eq5.44}
U_{screen}=\frac{1}{2}\sum_{i=1}^{N}\sum_{j=1}^{N}\sum_{|\bf{n}|=0}^{\infty}{'}\frac{{q_i}{q_j}}{4\pi\epsilon_{0}} \frac{{erfc}(\alpha|{\bf r}_{ij}+{\bf n}|)}{|r_{ij}+{\bf n}|}
\ee
This contribution is frequently referred to as `real space' summation
and its rate of convergence depends upon  the  parameter $\alpha$ of
the neutralising Gaussians. The larger the value of $\alpha$, the
narrower the gaussian distribution and the shorter the range of
interaction of the screened charges. Typically, $\alpha$ is chosen to
ensure that the minimum image convention can be applied i.e. the range
of the interactions is shorter than half the length of the simulation cell i.e.
only the $\vert {\bf n}\vert =0$ case needs to be considered. 
One can now rewrite:
\begin{eqnarray} 
U_{screen} &= &\frac{1}{2}\sum_{i=1}^{N}\sum_{j=1}^{N}\frac{{q_i}{q_j}}{4\pi\epsilon_{0}} \frac{{erfc}(\alpha|{\bf r}_{ij}|)}{|r_{ij}|}\\
\  &=& \frac{1}{2}\sum_i^N u_{i}^{screen}
\end{eqnarray}
where $u^{screen}_i$ represents the
contribution from the screened charges to the tagged atom Coulombic
energy of atom $i$.

The contribution of the Gaussian charge distribution, $\rho_{gauss}(\br )$, 
to the Coulombic interaction energy can be evaluated very simply in 
reciprocal or $k$-space and is denoted by $U_{rec}$.  
The electrostatic potential, $\Phi_{gauss}$, associated
with this charge distribution can be evaluated in real space using:
\be
-\nabla^2 \Phi_{gauss}(\br ) = 4\pi\rho_{gauss}(\br )
\ee
or in the Fourier form in reciprocal space as:
\be
k^2\Phi_{gauss}({\bf k}) =4\pi \rho_{gauss}({\bf k})
\ee
where $\rho_{gauss}({\bf k})$ is the Fourier transform of $\rho ({\bf r})$.
Fourier transforms of arrays of Gaussians can be computed analytically
and one can then write\cite{fs02}
\be
U_{rec}= \frac{1}{2}\sum_iq_i\Phi_{gauss}(\br )
\ee
Taking explicit summations over the partial charges, indexed by $j$,
 and the reciprocal lattice vectors, indexed by $\vert {\bf n}\vert$
one can write
\be
U_{rec} = \frac{1}{2 V_{0}\epsilon_{0}}\sum_{|n| \neq 0}^{\infty} \frac{\exp(-k^{2}/4\alpha^{2})}{k^2} \left | \sum_{j}^{N}q_{j} \exp(-i\vec{k}\cdot\vec{r}) \right |^{2}
\ee
Using the above expression, $U_{rec}$ can be calculated using a double loop over $\vert {\bf n}\vert$ and $j$. 
Computing the tagged atom contribution, requires the modulus squared term to be replaced by a 
double sum over $i$ and $j$, where $i$ and $j$ label the partial charges. 
Rearranging the above expression
\begin{eqnarray}
U_{rec} &  =  & \frac{1}{2V_{0}\epsilon}\sum_{j}\sum_{k\neq 0}\frac{q_j}{k^2}\exp(i{\vec k}\cdot{\vec r_{j}}-{k^{2}/4\alpha^{2}})
\sum_{i=0}^{N} q_{i} \exp(-i\vec{k}\cdot\vec{r_n}) \\
   &  =  &  \frac{1}{2V_{0}\epsilon}\sum_{j}\sum_{k\neq 0}\frac{q_j}{k^2}\exp(-{k^{2}/4\alpha^{2}})
\sum_{i=0}^{N} q_{i} \exp(-i\vec{k}(\vec{r_j}-\vec{r_i}) \\
        &  =  & \frac{1}{2} \sum_{j} u_{j}^{rec} 
\end{eqnarray}
Since the $u^{rec}_j$ contribution  multiplied by $(-i\vec{k})$ gives
 the force acting on charged site  $j$, the Ewald summation subroutines supplied
with the DL\_POLY program require only a small modification in order
to extract $u^{rec}_j$ \cite{sf96,syr01}.

It should be noted that the above expression for $U_{rec}$ includes a
self-interaction term for the
interaction of each Gaussian charge distribution with itself.  
When computing the Ewald summation 
to evaluate $U_{coul}$, it is necessary to remove these redundant 
self-interaction contributions, $U_{self}$,  which are given by
\be
U_{self}  = \sum_{j=1}^N u_{j}^{self} =(1/4\pi\epsilon )\sum_j 
\frac{q_j^2\alpha }{\sqrt{\pi }}
\ee
The complete expression for $U_{coul}$ will be 
\be
U_{coul}= U_{rec}+U_{screen}-U_{self}
\ee
 The contributions to the tagged potential energy that comes from the  
electrostatic interactions is given by 
$(u_{i}^{rec}+u_{i}^{screen}-u_{i}^{self})$.
The overall tagged molecule potential energy, $u_i$, of
molecule $i$ will therefore be given by :
\be
u_{i}=u_{i}^{vdw}+u_{i}^{rec}+u_{i}^{screen}-u_{i}^{self}
\ee

\section{Power Spectral Analysis}

The power spectral density of an observable $A(t)$ as a function of time $t$, 
has been  defined in equation (3). In this work, the dynamical observables are
the tagged atom potential energies, $u_i(t)$, where $i$ corresponds to a Si or
O atom.
Since we sample the potential energy signal at discrete times, the definition of the power spectral density must be changed to allow for discrete sampling \cite{numrec}:
\be
S_k = S(f_k) = \frac{\left| F_k \right|^2}{N^2}
\ee
where $N$ is the number of samples, and $F_k$ is the Discrete Fourier Transform (DFT) evaluated at the $k$-th frequency, i.e., 
\be
F_k = F(f_k) = \sum_{n=0}^{N-1} A_n e^{2\pi i n k/N}
\ee
(here we use the same normalization as ref. 20, the $k$-th frequency is $f_k = k/(N\Delta t)$, $\Delta t$ is the sampling interval, and  $k=0, \dots, N-1)$.
The dynamical time scales present in the system will clearly determine
the behaviour of the power spectrum. 
Consider a simple situation in which there is a single relaxation frequency
 $\lambda$ in the system. If  several such relaxation events take place at
an average rate $n$ at times $t_k, k=1, 2, \dots$, then each such event 
will give rise 
to an exponential decay in some appropriate dynamical variable of the
form $A_0\exp [-\lambda (t-t_k)]$ for $t\geq t_k$ (and 0
for $t\leq t_k$) leading to an overall time-dependent 
signal $A(t)=\sum_k A(t,t_k)$. The power spectrum of the fluctuating part of this signal is \cite{em02}
\be
S(f) = \frac{A_0^2n}{\lambda^2 + (2\pi f)^2}
\ee
which is frequently referred to as Debye relaxation.  In the limit 
$f \gg \lambda $, the Debye term approaches a simple $1/f^2$  behaviour.

If there are multiple time-scales, i.e. if the relaxation frequency
 of each of the single events is drawn from a probability distribution, 
$g_\lambda (\lambda )$, then the power spectrum must be computed as:
\be
\label{psd2}
S(f) = \frac{n \langle A^2 \rangle}{2\pi} \int_{\lambda_{min}}^{\lambda_{max}} \frac{g_\lambda(\lambda)}{(2\pi f)^2 + \lambda^2} d\lambda
\ee
If $g_\lambda (\lambda )$ corresponds to a uniform distribution  over the
range $\lambda_1$ to $\lambda_2$ and the amplitude of each pulse remains 
constant, then
\begin{eqnarray}
S(f) &=& \frac{1}{\lambda_2-\lambda_1}\int_{\lambda_1}^{\lambda_2 }
\frac{A_0^2 n}{\lambda^2 + (2\pi f)^2} d\lambda \nonumber \\
&=&\frac{A_0^2 n}{2\pi f(\lambda_2-\lambda_1)}\left[\arctan (\lambda_2/2\pi f )
-\arctan (\lambda_1 /2\pi f )\right]
\end{eqnarray}
In the low-frequency limit such that $f \ll \lambda_1\ll \lambda_2$, the spectrum flattens and we observe white noise
with $\langle S(f) \rangle= A_0^2 n$. In the intermediate region
where $\lambda_1 \ll f \ll \lambda_2$, we obtain 
\be
S(f) =\frac{A_0^2 n}{4 f(\lambda_2-\lambda_1)}
\ee
which corresponds to the \onebyf\ regime with $\alpha =1$. 
In the high-frequency limit for which $f \gg \lambda_2 \gg \lambda_1$, we obtain a $1/f^2$ noise with $S(f)= A_0^2n/(2\pi f)^2$. 
Other distributions
of time scales may also give rise to \onebyf\  behaviour; for example, 
a distribution $P(\lambda ) \propto \lambda^{-\beta}$ over the same
frequency interval will generate $S(f) \propto 1/f^{1+\beta}$.
Ref. \cite{em05} provides a gallery of spectral densities associated with
 different models.
In general, a spectral index $\alpha$ in the interval $0.5 < \alpha < 1.8$ may be taken as
an indication of multiple time-scale dynamics while exponents very close to
2 indicate single time-scale behaviour. 
A separation of two  decades
in time scales between $\lambda_1$ and $\lambda_2$ is sufficient 
to observe a \onebyf\ regime with $\alpha$ close to 1. 
Unless otherwise
stated, in this work, reference to $1/f^\alpha$ behaviour will imply the
multiple time-scale regime with $\alpha$ between 0.5 and 1.8, 
rather than the Debye tail with $\alpha \approx 2$.

In the case of a networked liquid, like water, the multiple time scale
region spans approximately two decades. There may be additional features,
such as high-frequency resonances or low-frequency Debye relations. One can,
in principle, disentangle the different contributions to the complex 
network dynamics using a model-based fitting procedure \cite{em05}. 
For the purpose of quantitatively characterising the
changes in the hydrogen bond network dynamics in the diffusionally
anomalous region, we have found it sufficient to identify the $1/f^\alpha$
region from the $\ln S(f)$ versus $\ln f$ plot and obtain the scaling 
exponent $\alpha$ by numerical fitting.

\section{Computational Details}

\subsection{Potential Model}

We use the van Beest-Kramer-van Santen (BKS) potential 
to model interatomic interactions in silica since it has been
extensively used to study liquid state anomalies and the glass transition 
\cite{bks90,kfb91}. The BKS potential is pair-additive and contains long-range
Coulombic and short-range two-body contributions.
The pair interaction between atoms $i$ and $j$ is given by:
\be
\phi_{BKS}(r_{ij}) = \frac{q_iq_j}{4\pi\varepsilon_0r_{ij}} + 
A_{ij}exp^{-b_{ij}r_{ij}} - \frac{C_{ij}}{r^6_{ij}}
\ee
where $r_{ij}$ is the distance between atoms $i$ and $j$ carrying charges
$q_i$ and $q_j$ and $A_{ij}$, $b_{ij}$ and $C_{ij}$ are the parameters
associated with the Buckingham potential for short-range repulsion-dispersion
interactions.  In the original BKS potential, the Buckingham parameters for 
the Si-Si interactions are taken as zero.  At high temperatures and pressures, 
the BKS potential exhibits an unphysical divergences in the interaction energy when the
Si-O distance becomes very small.  To remove these divergences, 
a short range correction term was added by Poole et. al \cite{vsp00} to 
the BKS potential and  the parameters adjusted in such a way that the 
original form was left unchanged at larger separations while 
the negative divergence was avoided at smaller separations. The modified 
pair interaction is
\be
\phi(r_{ij}) = \phi_{BKS} + 4\epsilon_{ij}\left[\left(\frac{\sigma_{ij}}{r_{ij}}\right)^{30} - \left(\frac{\sigma_{ij}}{r_{ij}}\right)^6\right]
\ee
where $\epsilon_{ij}$ and $\sigma_{ij}$ are the energy and length scale
parameters for the 30-6 Lennard-Jones interaction. The parameters for the
modified BKS potential used in this work are given in Table I.

\subsection{Molecular Dynamics}

Molecular Dynamics simulations of a system of 150 Si and 300 O ions were 
carried out in canonical ({\it N-V-T}) ensemble, using the DL\_POLY 
software package \cite{sf96,syr01}, under cubic periodic boundary conditions.
The effects of electrostatic (long-range) interactions were accounted for by 
the Ewald summation method \cite{at86,fs02}. The non-coulombic part was
 truncated and shifted at 7.5${\mathring{A}}$. A Berendsen thermostat, 
with the  time constant $\tau_{B}$=200ps, was used to maintain the 
desired temperature for the production run. The Verlet algorithm
with a time step of 1fs was used to integrate the equations
of motion.  The system was simulated at 5 temperatures
in the 4000K to 6000K range lying along 8 isochores  in the
density range from 1.8 g cm$^{-3}$ to 4.2 g cm$^{-3}$. 
For temperatures above 5000K, the system was equilibrated for 3-4ns 
followed by a production run of 5-7ns. For lower 
temperatures,  an equilibration period of 5-6ns was followed by a 
production run of 8-10ns. The starting configuration for the
liquid state simulations was generated by replicating the cubic unit cell 
of $\beta$-cristobalite to generate a cube of 
edge length 21.48${\mathring{A}}$ \cite{pear,wyck} which was 
then suitably truncated to
obtain a simulation cell containing 150 Si and 300 oxygen atoms.
The system thus obtained was heated gradually to form the equilibriated
liquid state.  Our simulation results are in 
agreement with previous results\cite{sdp02,vsp00} within the statistical 
error bars.

Figure[1] shows the $P({\rho})$ curve at different temperatures. It can be seen 
that at all densities below 3.0 g cm$^{-3}$, the pressures are negative 
for the temperatures studied, indicating that liquid silica is stretched 
rather than compressed. The isothermal compressibility, defined 
as  $\kappa_T =(1/\rho )(\partial \rho/\partial P )$, is negative only 
for the $\rho =1.8$ g cm$^{-3}$ isochore.
The spindonal density for which $\kappa_T =0$ is located around 2.0 g cm$^{-3}$.
 Below this density, the system is likely to be
inhomogenous. In our simulations,  the state points along the 1.8
 g cm$^{-3}$ isochore are thermodynamically unstable; however, to present 
a complete study to explore the diffusional anomaly we consider this
density.

\subsection{Power spectral analysis}
 
In this study, we focus exclusively on the power spectra associated with
tagged atom potential quantities, $u_O$ and $u_{Si}$, which were stored
at intervals of 10 fs during the MD runs. 
We represent power spectra of tagged potential energy fluctutations as
$S_u(f)$ but while referring to $S_u(f)$ spectra of atoms of a specific
type A, we use the notation $S_A(f)$.  The sampling interval of 10 fs corresponds
to a Nyquist frequency of 1666 cm$^{-1}$. The value of  the Berendsen thermostat
time constant ($\tau_B$)  provides
the lower limit on the frequency range over which we can obtain reliable power
spectra; thus, $\tau_B=200$ps corresponds to a lower frequency limit of 0.165
cm$^{-1}$.  Standard Fast Fourier Transform routines were used with
a square sampling window \cite{numrec}. The normalisation convention
was chosen such that the integrated area under the $S(f)$ curve
equalled the mean square amplitude of the time signal.
Windows containing $2^{19}$ data points were used for Fourier transformation.
Statistical noise in the power spectra was reduced by averaging over overlapping time signal windows as well as over individual tagged particle spectra.
In a given frequency interval showing \onebyf\ behaviour,
linear least squares fitting of $\ln S(f)$ was
done to obtain the $\alpha$ values.

\section{Results}

Figure 2 shows the behaviour of the self-diffusivities, $D_{Si}$ and $D_{O}$ 
of Si and O atoms respectively,
in BKS silica melt as a function of  density along different isotherms.
At any given state point, the oxygen self-diffusivities are  
greater than those of silicon atoms by approximately a factor of two
but the density and temperature-dependent trends are very similar.
Subsequently, in this paper, we will focus only on $D_{Si}$ since
silicon atoms are the sites with the local tetrahedral symmetry.
The diffusionally anomalous region is clearly demarcated in the
isotherms lying between 4000K and 5000K. The density of minimum
diffusivity, $\rho_{min}$, is approximately 2.0 g cm$^{-3}$ over this
 temperature range while the density of maximum diffusivity, 
$\rho_{max}$, is approximately
3.4 g cm$^{-3}$. In the case of SPC/E water, the region of anomalous 
diffusivity is seen below 280K and the $\rho_{min}$ and $\rho_{max}$ values
are approximately 0.9 g cm$^{-3}$ and 1.1 g cm$^{-3}$ respectively.

Figure 3 compares the power spectra generated by fluctuations in
potential energies of tagged oxygen and silicon atoms in 
silica melt at 4000K along the 2.0 g cm$^{-3}$ isochore with that of
water at 0.9 g cm$^{-3}$ at 230K.  
The chosen state point for both systems corresponds to
the density minimum along the appropriate isotherm. Despite the difference in 
intermolecular interactions between a molecular liquid and an ionic melt, 
there is a striking parallel in the power spectral features suggesting a
very similar pattern in the dynamical correlations. 

The $S_{Si}(f)$ spectrum  shows a sharp 
peak at 1100 cm$^{-1}$ corresponding to the  to the Si-O-Si 
asymmetric stretch vibration seen in both the experimental as well as 
simulated IR spectrum \cite{hk03,kyp06}. The power spectrum also shows a secondary, 
less intense peak at 800 cm$^{-1}$ corresponding to O-Si-O bending. 
It should be noted that the tagged particle power spectrum reflects
 the intrinsic dynamics of the liquid state network while the infra-red 
spectrum highlights those modes which couple to the electric field of the 
electromagnetic radiation in the IR frequency range. 
The localised high frequency peaks at 1100 and 800 cm$^{-1}$ correspond to 
well-defined local vibrational modes of the SiO$_4$ tetrahedra and  
mirror the librational peak in water which is, however, somewhat broader and
more intense.  

The region between 0.2 and 400 cm$^{-1}$ in $S_{Si}(f)$ may
be regarded as essentially a multiple time scale or $1/f^\alpha$ region
with a frequency-dependent value of $\alpha$.  At higher temperatures or
densities, the very low frequency region shows a cross-over to white noise.
The multiple time scale region of water spans a similar frequency range.

The $S_{Si}(f)$ and $S_{O}(f)$ spectra are qualitatively very similar. 
The tagged oxygen power spectrum,
$S_O(f)$, typically has a lower intensity than the $S_{Si}(f)$ power spectrum
because fluctuations in the tagged energy of the oxygen atoms are
 significantly lower than those for the Si atoms due to the relative
 magnitude of the charges. 
The high-frequency peaks in the $S_O(f)$ spectra occur at essentially the
same locations as in $S_{Si}(f)$ but have significantly different intensities
suggesting that tagged energies of the two types of atoms have different
sensitivities to the local vibrational modes.

Section 2 discusses the different contributions to the tagged atom
potential energy i.e. van der Waals ($u_{vdw}$), 
screened charge ($u_{screen}$), self-interaction ($u_{self}$)
 and reciprocal space ($u_{rec}$). While $u_{self}$ is a constant
for a given atomic type, the power spectra associated with fluctuations
in  the other three quantities at the state point (T=4000K, 2.0 g cm$^{-3}$)
is shown in Figure 4.  The behaviour of $S_u(f)$ is almost exactly reproduced 
over much of the frequency range by the screened charge contribution.
The spectral power associated with the van der Waals and reciprocal
space contributions is much lower and therefore has a minor effect
on the overall spectrum except in the high-frequency region. 
Unlike in the case of water, the $u_{vdw}$ contribution produces a 
relatively unstructured power spectrum, specially in the case of oxygen, 
while the Coulombic interactions are responsible for much of the
high-frequency, short length scale structure. This is
corroborated by the static distributions of the various
contributions to the tagged particle potential energies shown in Figure
5. The width of the $P(u_{rec})$ and $P(u_{vdw})$ distributions is
much narrower than that of $P(u_{screen})$ both in the case of
Si and O atoms.

To understand the effect of pressure and temperature on the tetrahedral
network in liquid silica, we examined the power spectra as well as
static distributions of tagged particle potential energies as
a function of temperature and pressure. Figures 6 and 7 show the results
along the 2.0 g cm$^{-3}$ isochore and the 4000K isotherm. 

Figure 6(a) shows that, along a given isochore, the crossover to white noise
occurs at increasingly higher frequencies as temperature increases. 
Moreover the local vibrational peak
broadens and the upper bound of the frequency range for multiple time-scale
or $1/f^\alpha$ behaviour increases. This progressive coupling of local modes 
into network vibrations leads to a change in the exponent $\alpha$ of the
$1/f^\alpha$ region. At the lowest temperatures, the frequency dependence
of $\alpha$ is more pronounced. For example, at 4000K and 2.0 g cm$^{-3}$ one
can identify a high frequency regime from 20-200 cm$^{-1}$ with $\alpha =0.8029$
while between 1-20 cm$^{-1}$, the $\alpha$ value is 1.6526. Since we wish to use
$\alpha$ as a measure of the degree of coupling of local and network modes,
we consider $\alpha$ in the 20-200 cm$^{-1}$ region.
As we show below, this change is quantitatively correlated
with increasing frequency in the anomalous region. 

The effect of
increasing density or pressure on the power spectrum is shown in Figure 6(b).
The crossover to white noise occurs at about 1 cm$^{-1}$ for 
densities of 3.0 g cm$^{-3}$ or more. The effect of increasing pressure
results in loss of the bimodal
character of the high-frequency peak. The peak at approximately 800 cm$^{-1}$
corresponding to the O-Si-O bend is clearly much more sensitive to
pressure than the Si-O-Si asymmetric stretch; this is not surprising since
it is known that compaction results in distortion of bond angles
resulting in more efficient packing of SiO$_4$ tetrahedra.
This is different from the loss of the distinct identity of the librational
peak seen in water with increasing density or temperature.

Figure 7 shows the static distributions, $P(u)$ of the tagged atom potential
energies. At low temperatures or low densities, the $P(u_{Si})$ distributions
are multimodal with six distinct peaks. This suggests that the energetic
environment of a given Si atom depends on bond lengths and bond angles 
associated with the nearest four oxygen atoms forming the local SiO$_4$
tetrahedral unit.  When all the angles and bond lengths are optimal, the lowest
energy environment is obtained. The presence of distinct peaks suggests
that there is a small energetic barrier to transit from one energetic
environment to another. The effect of temperature is to allow for more
facile transitions between different energetic environments which results
in a loss of multimodality but no additional broadening of the distributions.
In contrast, the effect of pressure is to both broaden as well as destructure
the distributions. The $P(u_O)$ has two major and one minor peak at low
temperatures or low densities. With increasing temperature, the 
minor peak becomes part of the shoulder of the low energy peak. With
increasing density, the minor, central peak disappears and the overall
wiidth of the distribution broadens.
 
The earlier study on water has established the strong correlation between
the diffusivity and the scaling exponent $\alpha$ of the multiple time-scale
regime in the anomalous regime.  Figures 8 and 9 show that this relationship 
also holds in the case of liquid silica. 
Figure 8 shows the behaviour of the scaling exponent $\alpha_u^{Si}$ of
the $S_{Si}(f)$ spectra as a function of density along different isotherms.
The parallel with the diffusivity plot in Figure 2 is obvious. It should
be noted, however, that $\alpha$ is essentially constant at 1.5 at high
temperatures and densities when the liquid is in the
diffusionally normal regime.  Figure 9 shows a correlation plot 
of $\alpha_u^{Si}$ with $\ln D_{Si}$ where the clustering of $\alpha$
values around 1.5 can be seen.  

\section{Discussion and Conclusions}

This study demonstrates the strong parallels in the dynamical behaviour of the
liquid state networks in silica and water in the anomalous regime.
Power spectral analysis is a convenient tool for characterising the complex
dynamics of network-forming liquids, such as silica, since the
multiple time-scale behaviour of the network is captured by the range
and exponent of the $1/f^{\alpha}$ regime. It is particularly useful for 
understanding diffusion in the random network of the liquid.
 In normal liquids, diffusion
proceeds via collisions and relaxations of the local nearest-neighbour cage.
In networked liquids, local vibrational modes can couple to network
reorganisations, thereby facilitating diffusion. The details of the
process in water and silica differ, but in both cases, the degree of
coupling can be indexed by the exponent $\alpha$ of the multiple time-scale
regime. As a consequence, the diffusivity is strongly correlated 
with the scaling exponent, specially in the anomalous regime. As the network
connectivity is attenuated by temperature and pressure in the normal
regime, the correlation begins to break down.

The results of this study provide some insights into the microscopic
connection between the structural and diffusional anomalies. The
boundaries of the structurally anomalous region are determined by
extrema in the local order parameters. The progressive destruction 
of local tetrahedral order by temperature or pressure has an effect on
the local atomic or molecular environment, as can be seen from the
static distribution of the tagged particle potential energies.  As the local
tetrahedral order is destroyed, the vibrational coupling of local modes,
specially the O-Si-O bend, to the network reorganisations  becomes
much stronger causing a rise in diffusivity. The diffusivity does not,
however, rise indefinitely with increasing compression because for
sufficiently large values of the density, the extent and connectivity of
the network is severely attenuated and the liquid shifts from the anomalous
to the normal regime. Therefore one would expect the diffusional and 
structural anomalies to  be strongly correlated. Whether the structural
anomalies precede or succeed the diffusional anomaly (as in water and silica
respectively) must depend on the details of the potential
energy surface, such as the nature and extent of vibrational coupling. 
Based on the similarity in power spectral behaviour of SPC/E and TIP5P-E
models for water \cite{mcm06}, we would expect different effective
pair potentials for silica to have very similar qualitative behaviour.
It is possible, however,  that three-body potentials for silica
will show more significant deviations, such as a reversal in the
relative order of the two types of anomalies \cite{hk03}.

The parallel behaviour of power spectra of water and silica in the
diffusionally anomalous region suggests that all tetrahedral liquids should
display this generic pattern of behaviour.  It would be interesting 
to perform a power spectral analysis of other liquids showing water-like
anomalies, such as two-scale ramp potentials \cite{ybgs05}, to see if
the detailed dynamical behaviour resembles that of tetrahedral liquids.

{\bf Acknowledgements} This work was supported by the Department of Science 
and Technology, New Delhi,  under the Swarnajayanti Fellowship scheme. 
RS thanks Council for Scientific and Industrial Research,
New Delhi, for the award of a Junior Research Fellowship.

\newpage
{\bf Table I}
                                                                                
{BKS (modified) potential parameters \cite{bks90,kfb91}.\\
                                                                                
\bigskip
\begin{tabular}{cccccc}
\smallskip\\
\hline\hline
~&~&~&~&~\\
$i-j$  & $A_{ij}$  & $b_{ij}$  & $C_{ij}$  & $\epsilon_{ij}$  &
 $\sigma_{ij}$ \\
\   &  (kJ mol$^{-1}$) &  ($\mathring{A}^{-1}$) &  (kJ mol$^{-1}$) &  (kJ mol$^{-1}$) &  ($\mathring{A}$)\\
~&~&~&~&~\\
\hline
~&~&~&~&~\\
$O-O$  & 134015   & 2.76  & 16887.3 & 0.101425  & 1.7792 \\
$Si-O$ & 1737340  & 4.87  & 12886.3 & 0.298949  & 1.3136 \\
~&~&~&~&~\\
\hline\hline
\end{tabular}

\newpage
\begin{center}
{\bf Figure Captions}
\end{center}

\begin{enumerate}
\item Equation of state for BKS silica: Pressure ($P$) as a function of density ($\rho$) along different isotherms.
 Systems at a density of 1.8 g cm$^{-3}$ are likely to be inhomogeneous.

\item Dependence on density, $\rho$, of (a) self-diffusivity of Si, $D_{Si}$,
and (b) self-diffusivity of O, $D_{O}$ along different isotherms.

\item Comparison of power spectra associated with temporal fluctuations in 
tagged particle potential energy of Si abd O atoms at 4000K and 2.0 g cm$^{-3}$
with water at 230K and 0.9 g cm$^{-3}$. The power spectrum of O atoms has been 
scaled to half for clarity.

\item Contributions of reciprocal space and screening from Ewald summation 
and van der Waals interactions to the power spectra of total tagged 
potential energy fluctuations of (a) Si at 4000K and 2.0 g cm$^{-3}$
and (b) O at 4000K and 2.0 g cm$^{-3}$. Relative magnitudes kept unaltered.
  
\item Contributions of reciprocal space and screening from ewald summation
and van der Waals interactions to the static distribution of tagged particle 
potential energy of (a) Si at 4000K and 2.0 g cm$^{-3}$ and (b) O at 4000K and 2.0 g cm$^{-3}$. In part (a) van der Waals and reciprocal space contributions, 
P($u_{vdw}$) and P($u_{rec}$) respectively, have been scaled by a factor of 
$1/10$. In part (b) P($u_{vdw}$) has been scaled by $1/4$ while P($u_{rec}$)
has been scaled by $1/10$.

\item Power spectra of tagged particle potential energy fluctuations, $S_u(f)$,
of Si in liquid silica along (a) 2.0 g cm$^{-3}$ isochore from 4000K to 
6000K in steps of 500K and (b) 4000K isotherm at 1.8, 2.0, 2.2, 2.6, 3.0, 3.4,
 3.8 amd 4.2 g cm$^{-3}$.   
Relative magnitudes of power spectra for different 
temperatures and densities were adjusted (scaled) for clarity.

\item Static distributions of tagged particle potential energies
 (in kJ mol$^{-1}$) of (a) Si along 2.0 g cm$^{-3}$ isochore, (b) Si along
4000K isotherm, (c) O along 2.0 g cm$^{-3}$ isochore and (d) O along 
4000K isotherm.

\item Multiple time-scale exponent, $\alpha_u^{Si}$, as a function of density 
along different isotherms. The exponent $\alpha_u^{Si}$ is evaluated using the high frequency \onebyf\ regime of 20 - 200 cm$^{-1}$ 
at 4000K and  4500K and from the point of crossover to 200 cm$^{-1}$ 
for 5000K, 5500K and 6000K.

\item Correlation plot showing $\ln D_{Si}$ against $\alpha_u^{Si}$ for 
different isochores. The frequency range for evaluating
 $\alpha_u^{Si}$ are the same as in Figure 8. 
Units of diffusivity are taken as 10$^{-5}$ cm$^2$ s$^{-1}$.

\end{enumerate}

\begin{figure}
\centering
\includegraphics[trim= 2in 2in  0 0 ]{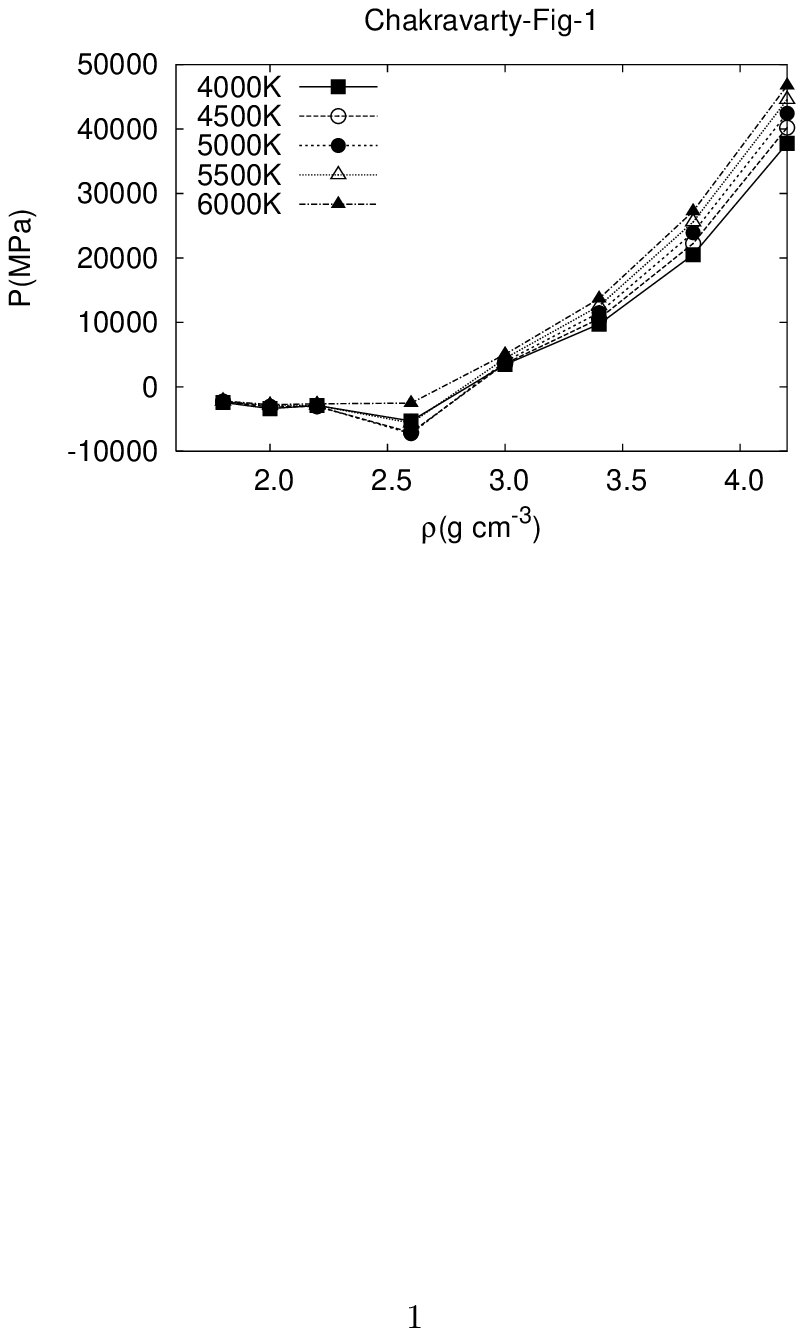}
\end{figure}
\newpage

\begin{figure}
\centering
\includegraphics[trim= 2in 2in 0 0]{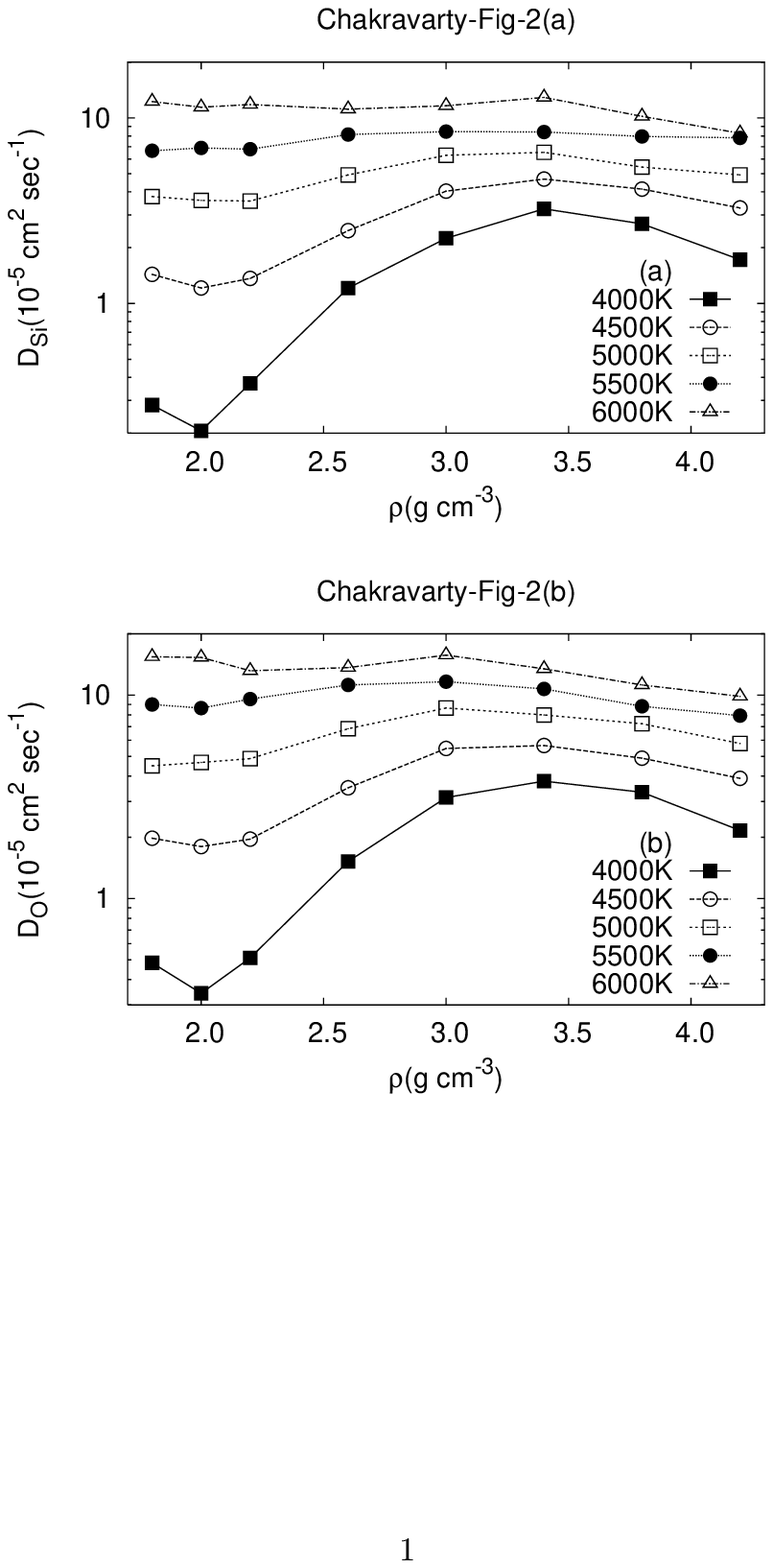}
\end{figure}
\newpage

\begin{figure}
\centering
\includegraphics[trim= 2in 2in 0 0]{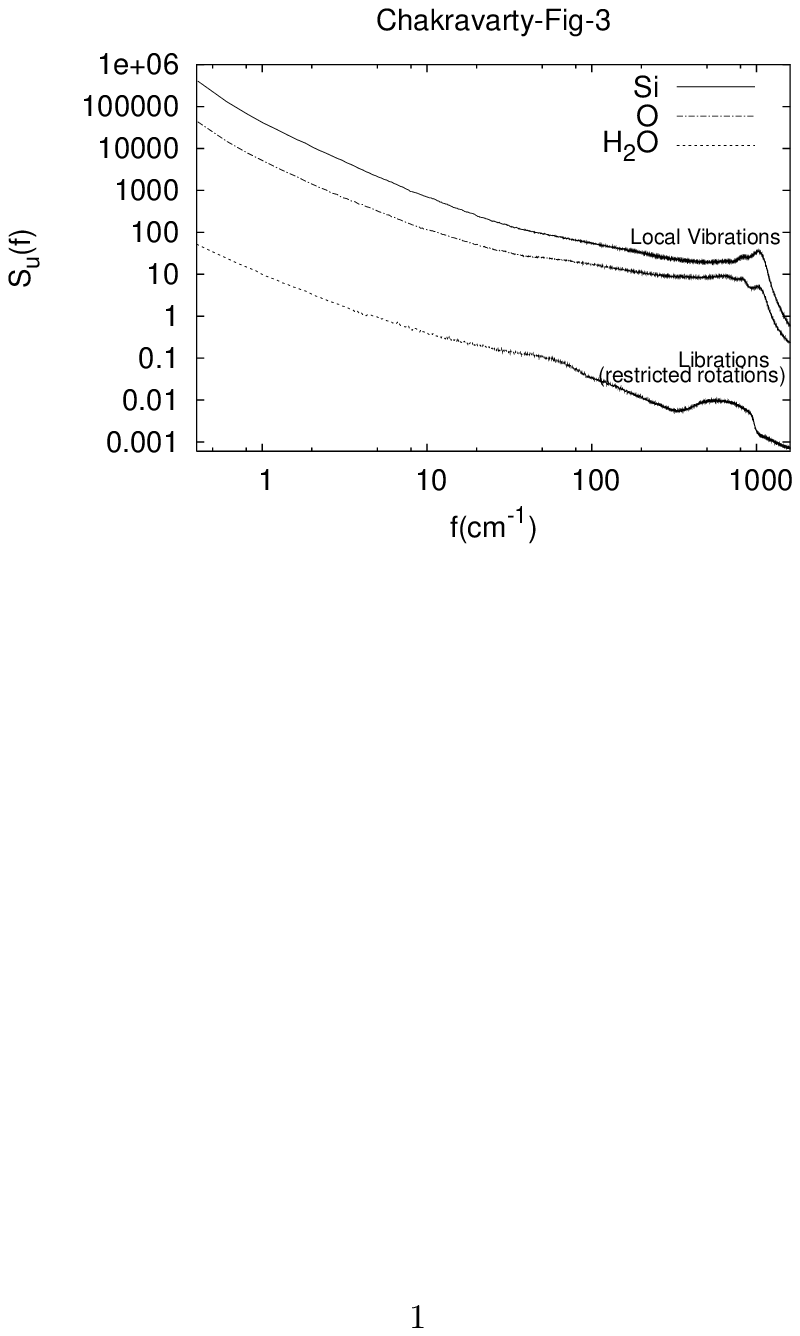}
\end{figure}
\newpage

\begin{figure}
\centering
\includegraphics[trim= 2in 2in 0 0]{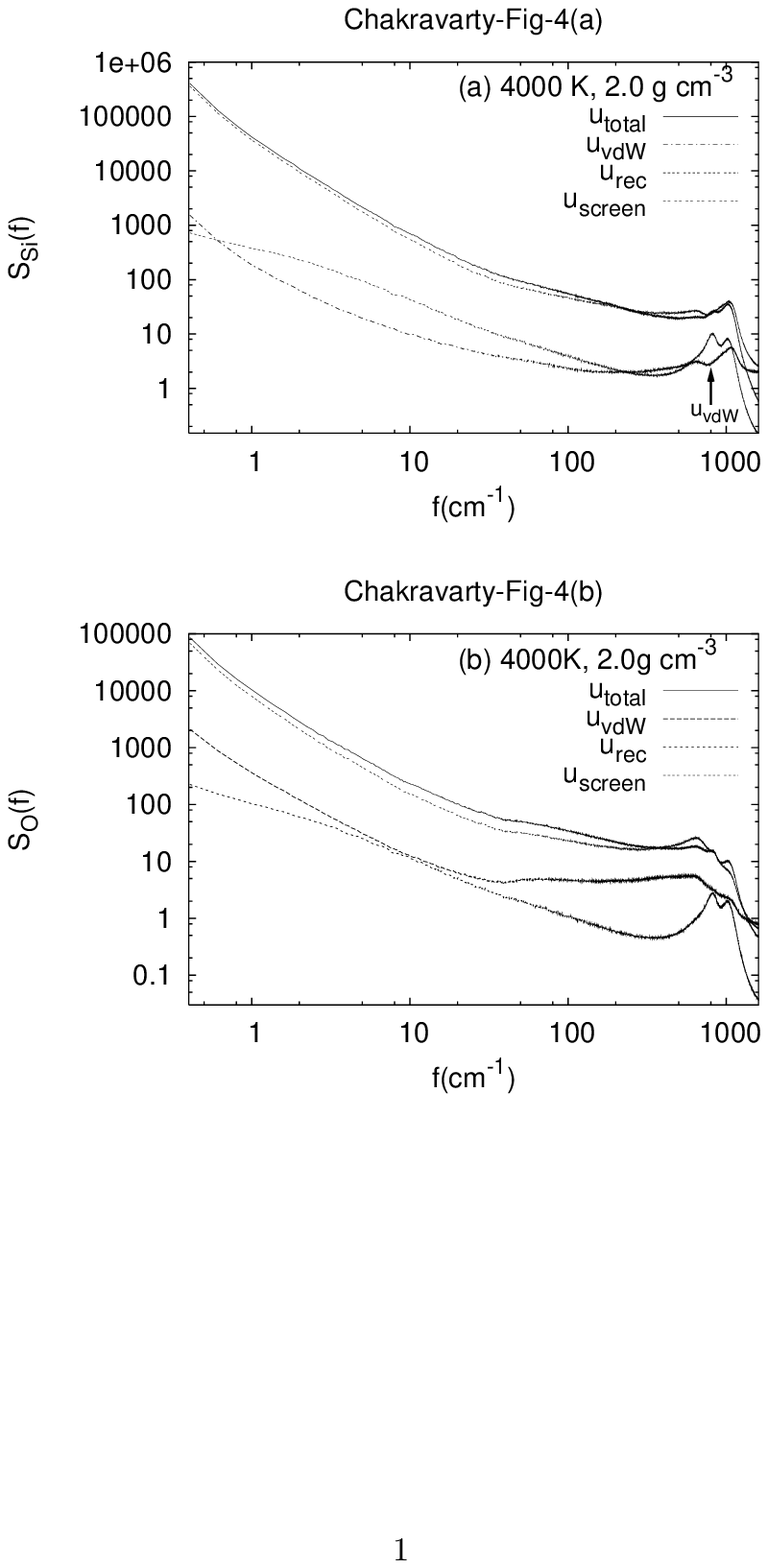}
\end{figure}
\newpage

\begin{figure}
\centering
\includegraphics[trim= 2in 2in 0 0]{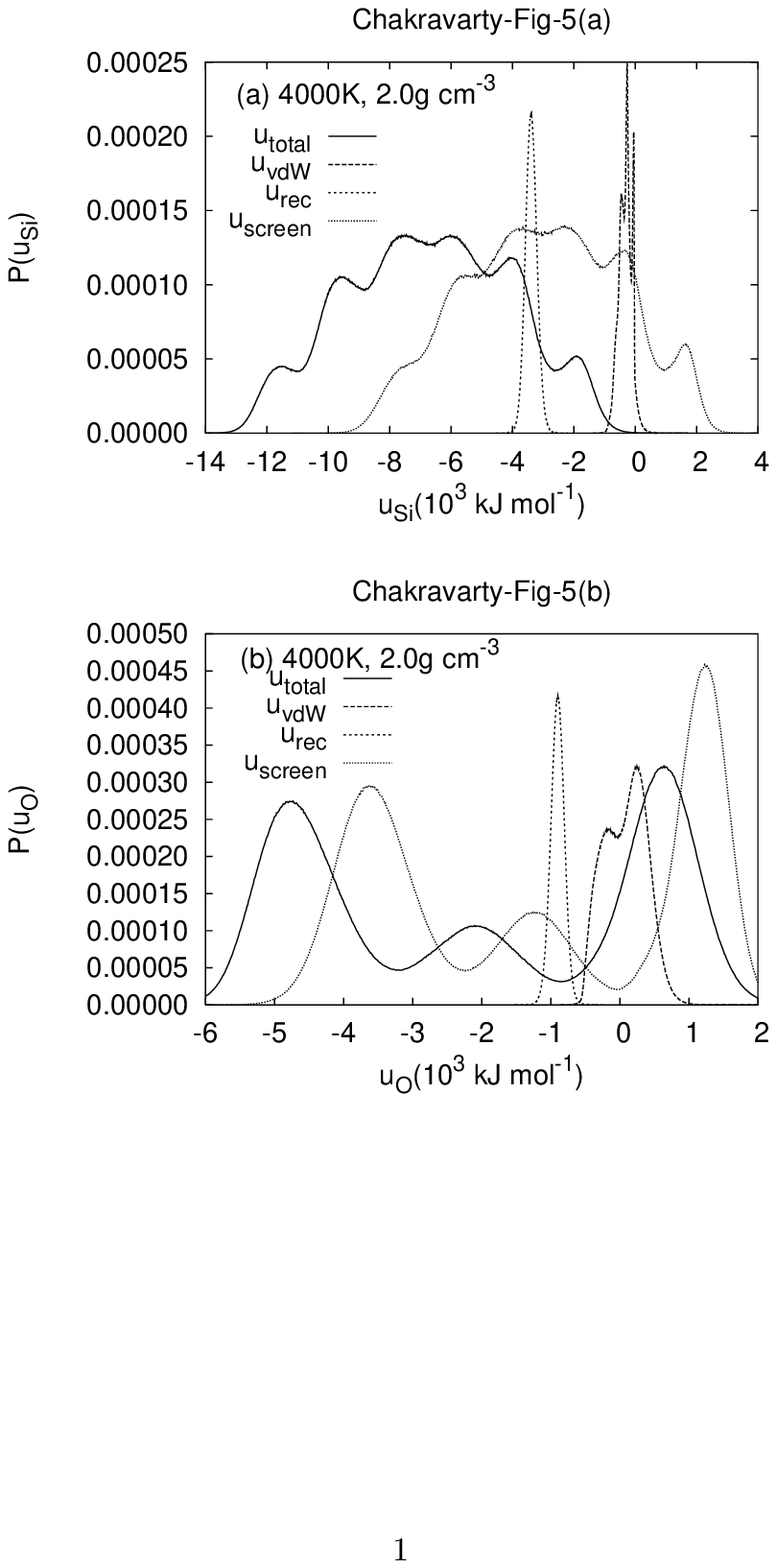}
\end{figure}
\newpage

\begin{figure}
\centering
\includegraphics[trim= 2in 2in 0 0]{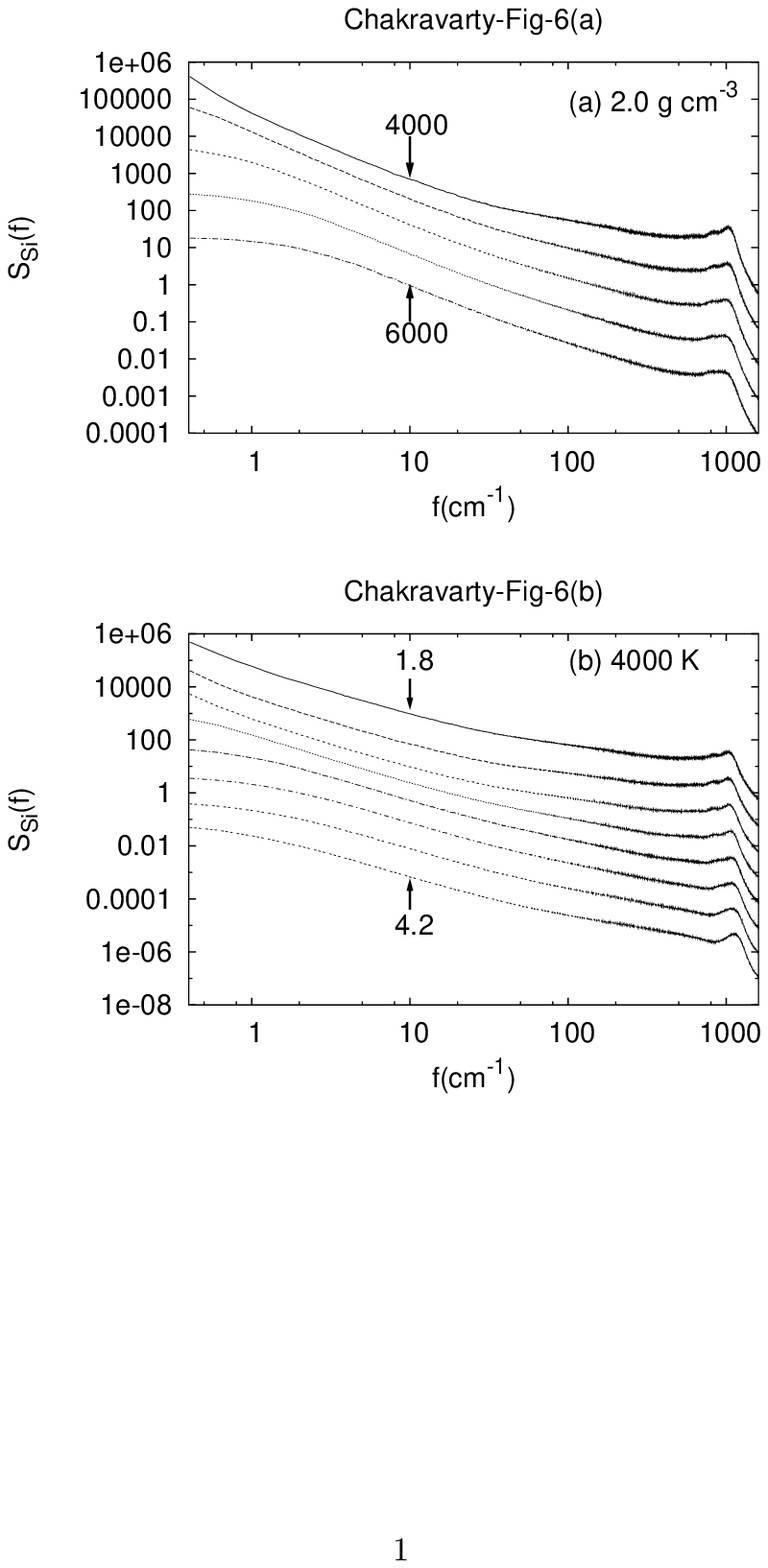}
\end{figure}
\newpage

\begin{figure}
\centering
\includegraphics[trim= 2in 2in 0 0]{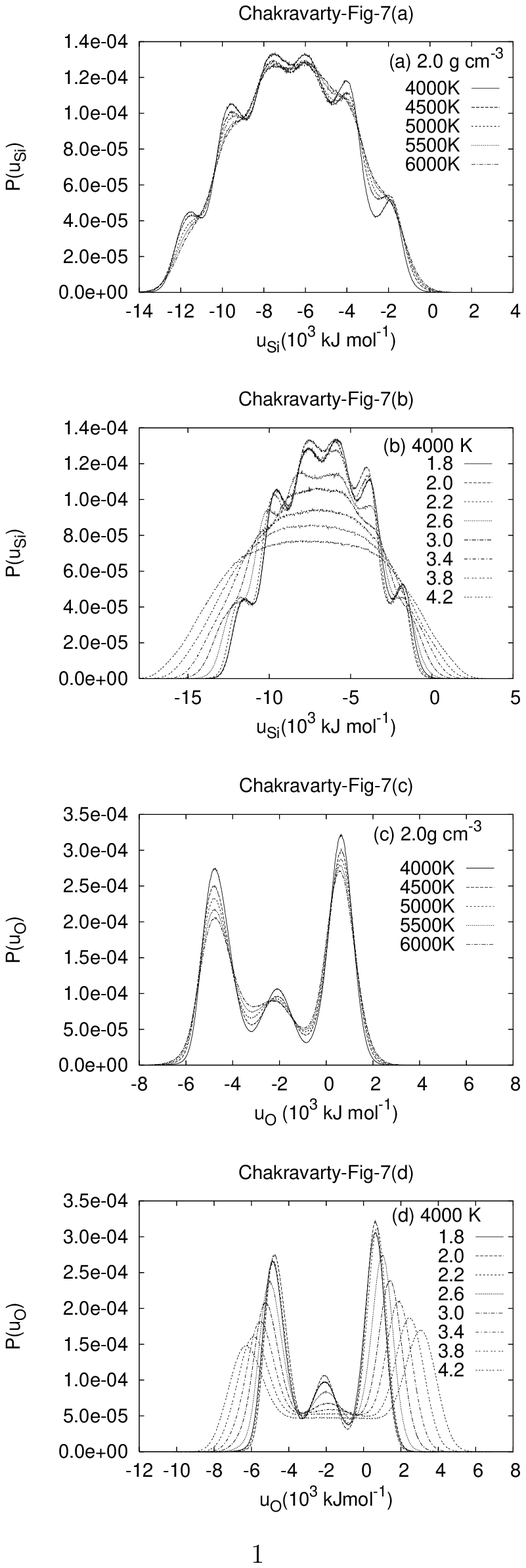}
\end{figure}
\newpage

\begin{figure}
\centering
\includegraphics[trim= 2in 2in 0 0]{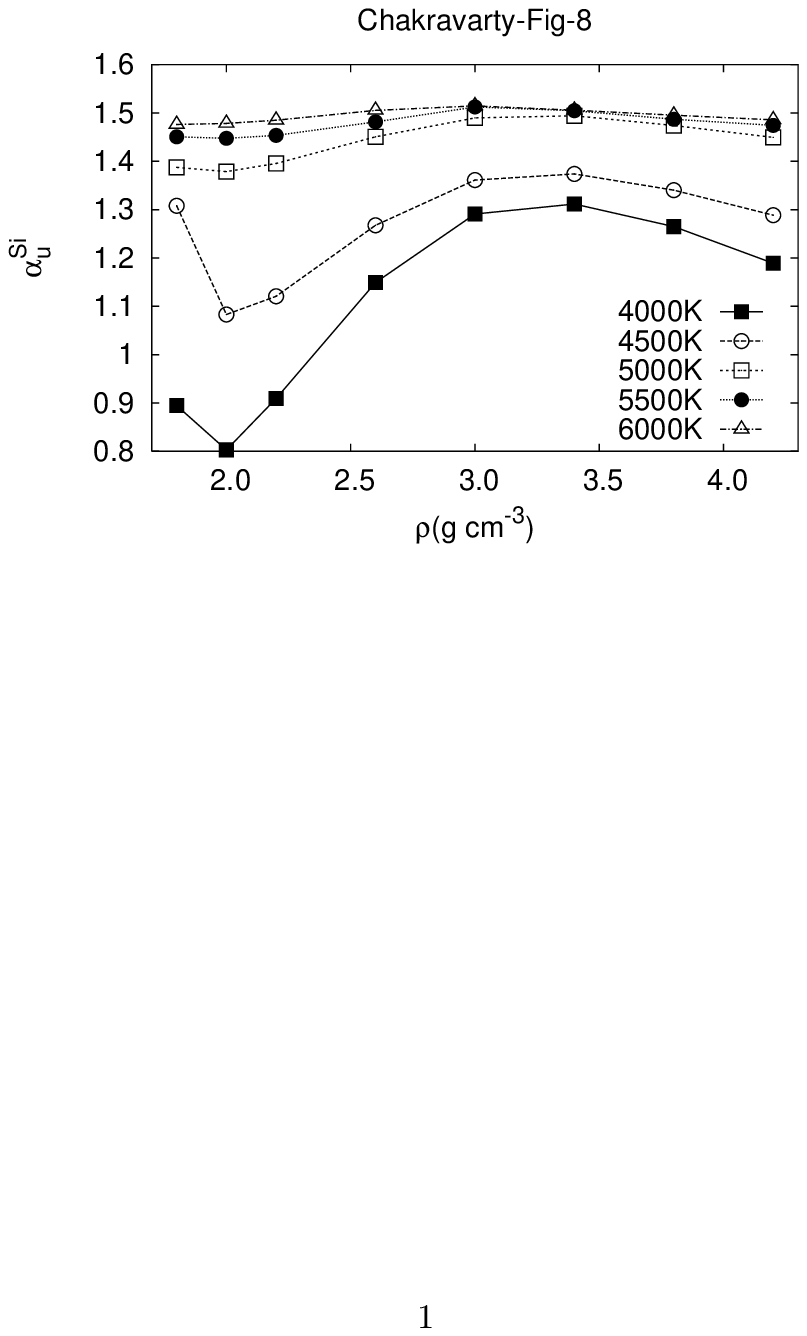}
\end{figure}

\begin{figure}
\centering
\includegraphics[trim= 2in 2in 0 0]{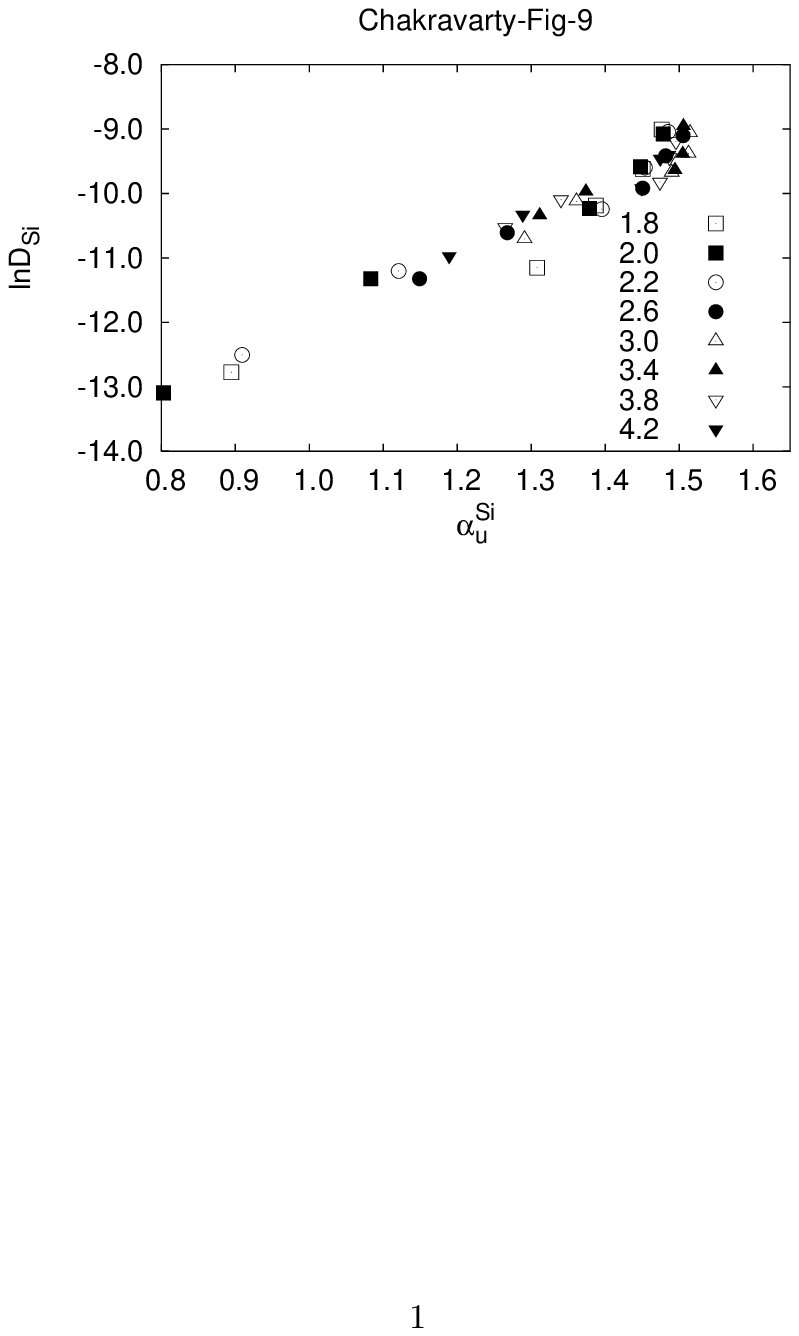}
\end{figure}

\end{document}